\documentclass[aps,prb,twocolumn,english]{revtex4-1}
\usepackage[latin9]{inputenc}
\usepackage{amstext,bm}
\usepackage{graphicx}
\usepackage{xcolor}
\usepackage{color}
\usepackage{amsmath}
\usepackage{babel}
\usepackage[titletoc]{appendix}
\usepackage{hyperref}
\hypersetup{
     colorlinks = true,
     citecolor  = red,  
     linkcolor  = magenta 
}
\usepackage{float}

\begin{document}

\title{Topological Node-Line Semimetal in Compressed Black Phosphorus}

\author{Jianzhou Zhao$^{1,4}$}
\author{Rui Yu$^{3}$}
\email{yurui@hit.edu.cn}
\author{Hongming Weng$^{1,2}$}
\email{hmweng@iphy.ac.cn}
\author{Zhong Fang$^{1,2}$}

\affiliation{$^{1}$
Beijing National Laboratory for Condensed Matter Physics, 
and Institute of Physics, Chinese Academy of Sciences, 
Beijing 100190, China}

\affiliation{$^{2}$
Collaborative Innovation Center of Quantum Matter, 
Beijing 100190, China}

\affiliation{ $^{3}$ 
Department of Physics, 
Harbin Institute of Technology, 
Harbin 150001, China}

\affiliation{ $^{4}$
Co-Innovation Center for New Energetic Materials,
Southwest University of Science and Technology,
Mianyang, Sichuan 621010, China}

\date{\today}
\begin{abstract}
Based on first-principles calculations and tight-binding model analysis,
we propose that black phosphorus (BP) can host a three-dimensional
topological node-line semimetal state under pressure when spin-orbit
coupling (SOC) is ignored.
A closed topological node line exists in the first Brillouin zone (BZ) 
near the Fermi energy, which is protected by the coexistence of time-reversal
and spatial inversion symmetry with band inversion driven by pressure.
Drumhead-like surface states have been obtained on the beard (100) surface.
Due to the weak intrinsic SOC of phosphorus atom, a band gap less
than 10 meV is opened along the node line in the presence of SOC and
the surface states are almost unaffected by SOC.
\end{abstract}


\maketitle

\section{Introduction}

The discovery of topological states of matter has attracted broad interests in recent years.
It was ignited by the discovery of two-dimensional (2D) and three-dimensional (3D) topological insulators (TIs)~\cite{Hasan:2010vc,Qi:2011wt,Bernevig_book,Weng:2014sva,Zhang:2012uw}.
These materials exhibit a bulk energy gap between the valence and conduction bands, 
similarly to normal insulators, but possess unique gapless
boundary states that are protected by the topology of bulk states.
Topological states has also been proposed for 3D metals with stable Fermi surface \cite{Horava_2005PRL}, 
as topological semimetals (TSM).
Recently, there have been great research interest for novel 3D metals with nontrivial 
band topology defined on their compact and continuous Fermi surface, 
which can be looked as the counterpart of the whole 2D BZ of insulator~\cite{Wang:2012ds}. 
Generally speaking, these 3D metals are topologically stable for their Fermi
surfaces enclose nontrivial degenerate band touch points (BTP). 
Such BTP behaves as the ``source" or ``sink" of Berry flux, and brings quantized Berry flux when passing
through the surrounding enclosed Fermi surface~\cite{Weng_ADV:2015hh}.
This quantized number can be taken as the topological invariant to identify the band topology of corresponding metals.
According to the Fermi-liquid theory, the physical properties of metals are mostly determined by the lower energy part around Fermi level. Therefore, the closer of the
Fermi level to the BTP, the better for manifesting the unique properties from the band topology of the Fermi surfaces. The ideal case is that the whole BZ has just the BTP points at the Fermi level.
In this ideal case, the density of states at Fermi energy is zero and that's why such nontrivial 3D metals are called as topological semimetals.
Up to now, three types of TSM, namely Weyl semimetal (WSM), Dirac semimetal (DSM) and node-line semimetal (NLS) based on the BTPs' degeneracy and distribution in BZ, have been proposed. 
For WSM, the BTPs are double degenerate, having definite chirality and appearing in isolated pairs, which are in fact the quasi-particle of Weyl fermions.
For DSM, the BTPs are fourfold degenerate and can be looked as the ``kiss" of two Weyl fermions with opposite chirality in BZ.
For NLS~\cite{Burkov:2011eg}, the BTPs around the Fermi level form closed loops in BZ. 
These three type of TSMs constitute the TSM ``trio''.
The intriguing expected properties characterizing above TSMs include 
the surface Fermi arc, the nearly flat surface bands, 
the negative magneto-resistivity due to chiral anomaly and the unique Landau energy level.

Presently, the experimentally extensively studied DSMs are Na$_3$Bi and Cd$_3$As$_2$,~\cite{Yang:2014ia} 
both of which were predicted theoretically~\cite{Wang:2012ds,Wang:2013is} and then confirmed by several 
experiments~\cite{PhysRevB.91.155139,PhysRevLett.113.246402,Neupane:2014kc,Liu:2014bf,Liu:2014hr}.
Starting from DSM, which serves as a singularity point of various topological quantum states, one can obtain 
WSM by breaking either time-reversal~\cite{Wan:2011tp,Xu:2011dy} or inversion symmetry~\cite{PhysRevB.85.035103,PhysRevB.90.155316}. 
The prediction of WSM state in nonmagnetic and noncentrosymmetric TaAs family~\cite{Weng:2015dy,Huang:2015ic} 
have been verified by experiments~\cite{Lv:2015pya,Huang:2015um,Lv:2015kp,Xu07082015,Xu:2015vb,Lv:2015vf}. 
For the third member of TSM, the coexistence of time-reversal and inversion symmetry protects NLS in 3D 
momentum space when SOC is neglected and band inversion happens~\cite{Weng:2014wg,Yu:2015wl,kim_NLS_2015PRL}.
Based on this, there have been several carbon allotropes 
re-examined or proposed as host of NLS, including Mackay-Terrones crystal (MTC) \cite{Weng:2014wg}, 
Bernal graphite~\cite{Heikkila:2015vj}, hyper-honeycomb lattices~\cite{Mullen:2014wq}, 
and the interpenetrated graphene network~\cite{Chen:2015vm}.
It was also shown that mirror symmetry, instead of inversion symmetry, together with time-reversal symmetry 
can protect NLS when SOC is neglected, like TaAs~\cite{Weng:2015dy} and Ca$_3$P$_2$~\cite{Xie:2015tq,chan_Ca3P2_NLS_2015}. 
Due to the correspondence between Dirac line in bulk and surface states at boundary
\cite{PhysRevLett.89.077002,Burkov:2011eg}, the characteristic surface ``drumhead''-like state is demonstrated on the NLS surface.~\cite{Weng:2014wg,Yu:2015wl,kim_NLS_2015PRL} 
Such 2D surface states with nearly zero dispersion are proposed as a route to achieving 
high-temperature superconductivity~\cite{PhysRevB.83.220503,Heikkila:2015vj}.

In the present work, based on first-principles calculations and 
tight-binding model Hamiltonian analysis,  we predict that BP 
under pressure is another candidate for NLS. The rest of the 
paper is organized as follows.
In section II, we present the crystal structure and the 
first-principles calculation methodology.
Then we present the bulk and surface electronic structure of
compressed BP form first-principles calculations 
in Sec.~\ref{sec_band}. In Sec.~\ref{sec_TBmodel}, a four-bands 
tight-binding model is constructed and the NL structure, 
the surface states related to the terminate way on the (100)
surface are studied from the four-bands tight-binding model.
Conclusions are given at the end of this paper.

\section{The crystal structure and computation Method}\label{sec_str}

\begin{figure}
\begin{centering}
\includegraphics[width=1\columnwidth]{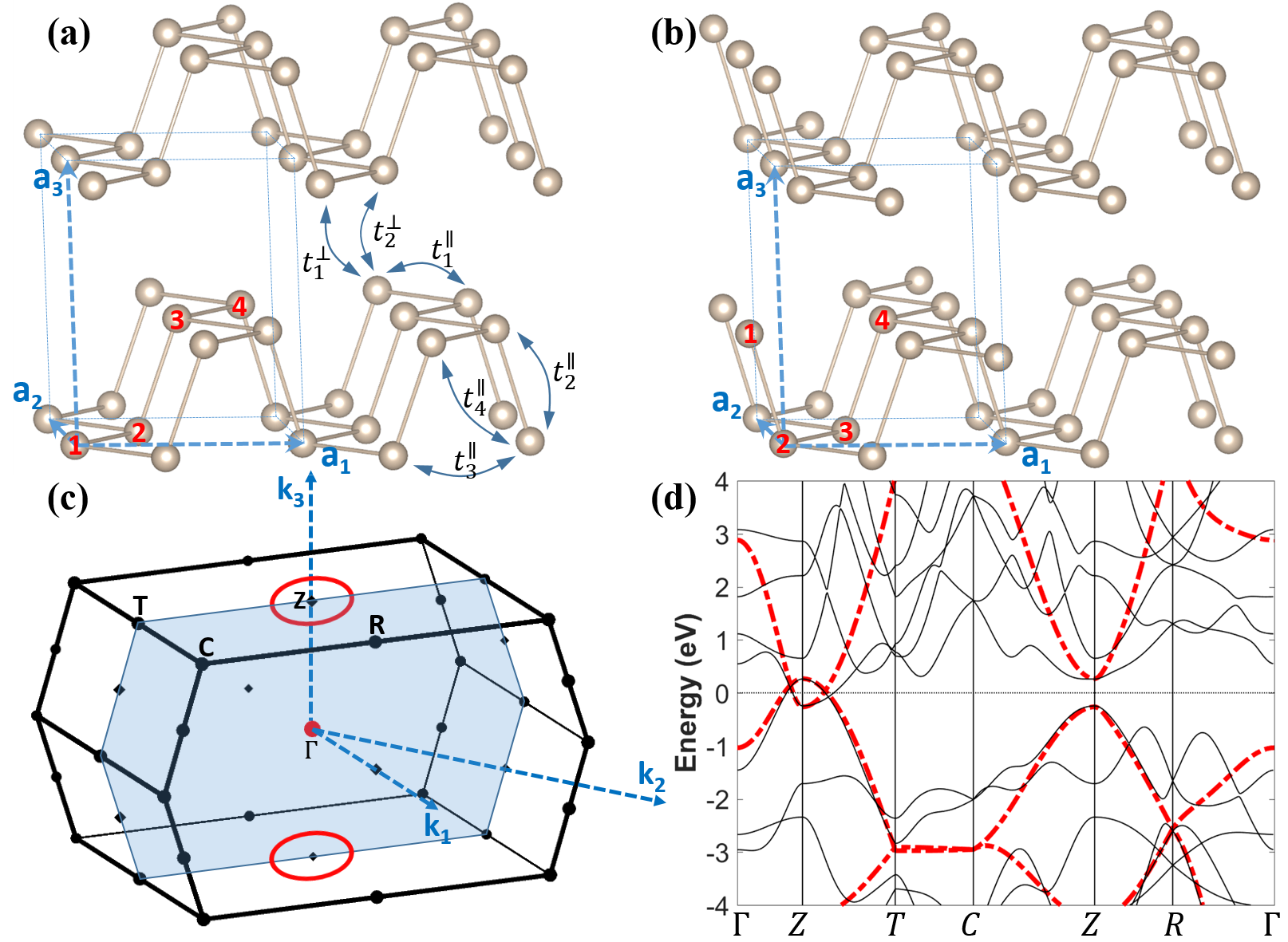}
\par\end{centering}
\protect\caption{\label{fig:CS_BZ}
(Color online)
(a) Crystal structure of bulk BP and the hopping parameters $t_{ij}$ of the tight-binding model. 
The surface in (100) direction with zigzag type and beard type
surface are presented in (a) and (b) respectively.
(c) Brillouin zone of the bulk BP. 
The node-line is schematically shown with red color circle
which surrounds Z point and lies on the T-Z-$\Gamma$ plane.
(d) The band structures of BP under hydrostatic pressure is shown 
with black color. Band structures calculated by using four-bands 
tight-binding model are shown with red dashed curves.}
\end{figure}

BP is the most stable allotrope of the element of phosphorus at ordinary temperatures and pressure, which was successfully synthesized 100 years ago \cite{doi:10.1021/ja02184a002} and attract great interest in its monolayers, few-layer and thin film structure in recent years \cite{Li:2014gf,Ling:2015ba,Tran:2014fb,Rudenko:2014je}.
It has been proposed that the properties of BP can be effectively controlled by strain \cite{Rodin:2014kx,Li:2014gfa,Peng:2014dp,Rodin:2014kx,Xiang:2015vo,Akiba:2015bi}.
A uniaxial compressive strain will switch BP from a direct band gap semiconductor to an indirect band gap semiconductor, semimetal, or metal \cite{Rodin:2014kx}.
A moderate hydrostatic pressure effectively suppresses the band gap and induces Lifshitz transition from semiconductor to semimetal accompany with SdH oscillations and giant magnetoresistance \cite{Xiang:2015vo,Akiba:2015bi}.
A semiconductor to band-inverted semimetal transition is also reported can be tunned by
a vertical electric field form dopants in few-layer BP and the DSM can be achieved at the critical field~\cite{Kim:2015di}.

The orthorhombic bulk BP belonging to the $Cmce$ space group (No. 64) with layered structure.
Each layer is a 2D hexagonal lattice which is puckered along the armchair direction. 
These monolayers are stacked along the z direction with van der Waals interactions between 
them as shown in Fig.~\ref{fig:CS_BZ}(a). 
In the following calculations, the primitive cell are used as illustrated in Fig.~\ref{fig:CS_BZ}.
In the ambient condition, the bulk BP is a semiconductor with a gap about 300 meV~\cite{Tran:2014fb}. 
Under quite low pressure, the energy of valence and conduction bands inverted, and the BP becomes semimetal~\cite{Fei:2015wf,Xiang:2015vo,Akiba:2015bi}.
In this work, we focus on the volume of $V\approx0.88V_{0}$ ($V_0$ is optimized equilibrium volume by GGA), which is in the band inversion region~\cite{Gong:2015wb} and corresponds to BP under 1.75 GPa pressure estimated from our calculation. We find that under such pressure, BP has robust band inversion and well defined semimetal
feature.
The first-principle calculations are performed by using the Vienna $ab\ initio$ simulation package (VASP) based on generalized gradient approximation in Perdew-Burke-Ernzerhof (PBE)~\cite{Perdew:1996iq} type and the projector augmented-wave (PAW) pseudo-potential~\cite{Blochl:1994zz}. The energy cutoff is set to 400 eV for the plane-wave basis and the BZ integration was performed on a regular mesh of 10$\times$14$\times$8
k-points. A tight-binding model based on maximally localized Wannier
functions (MLWF) method \cite{Mostofi:2008ff,Marzari:2012eu} has
been constructed in order to investigate the surface states in the 
(100) direction.

\section{Results and Discussion}

\subsection{Electronic structure}\label{sec_band}

The band structure of compressed bulk BP is presented in Fig.~\ref{fig:CS_BZ} (d).
Near the Fermi energy, two bands with opposite parity are inverted
at the Z point. The intriguing point of the band structure is that
the band crossings due to the band inversion form a  node ring
around the Z point in the T-Z-$\Gamma$ plane of the BZ as shown in Fig.~\ref{fig:CS_BZ} (c), which is protected by the coexistence of time-reversal and inversion symmetry as addressed in
Ref.~\onlinecite{Weng:2014wg,Yu:2015wl}.
Only one node line is presented in the BZ, which is different from the all-carbon Mackay-Terrones crystal (MTC)~\cite{Weng:2014wg} and antiperovskite Cu$_{3}$PdN~\cite{Yu:2015wl,kim_NLS_2015PRL} where three node lines exist due to the cubic symmetry in these systems.
In the present of SOC, gaps are opened along the node line.
It's about 10 meV along Z-$\Gamma$ direction and 2 meV along 
T-$\Gamma$ direction due to the weak SOC strength of phosphorus atoms.
These results show that the influence by SOC can be neglected as temperature higher than 100 K.

\begin{figure}[h]
\begin{centering}
\includegraphics[width=0.95\columnwidth]{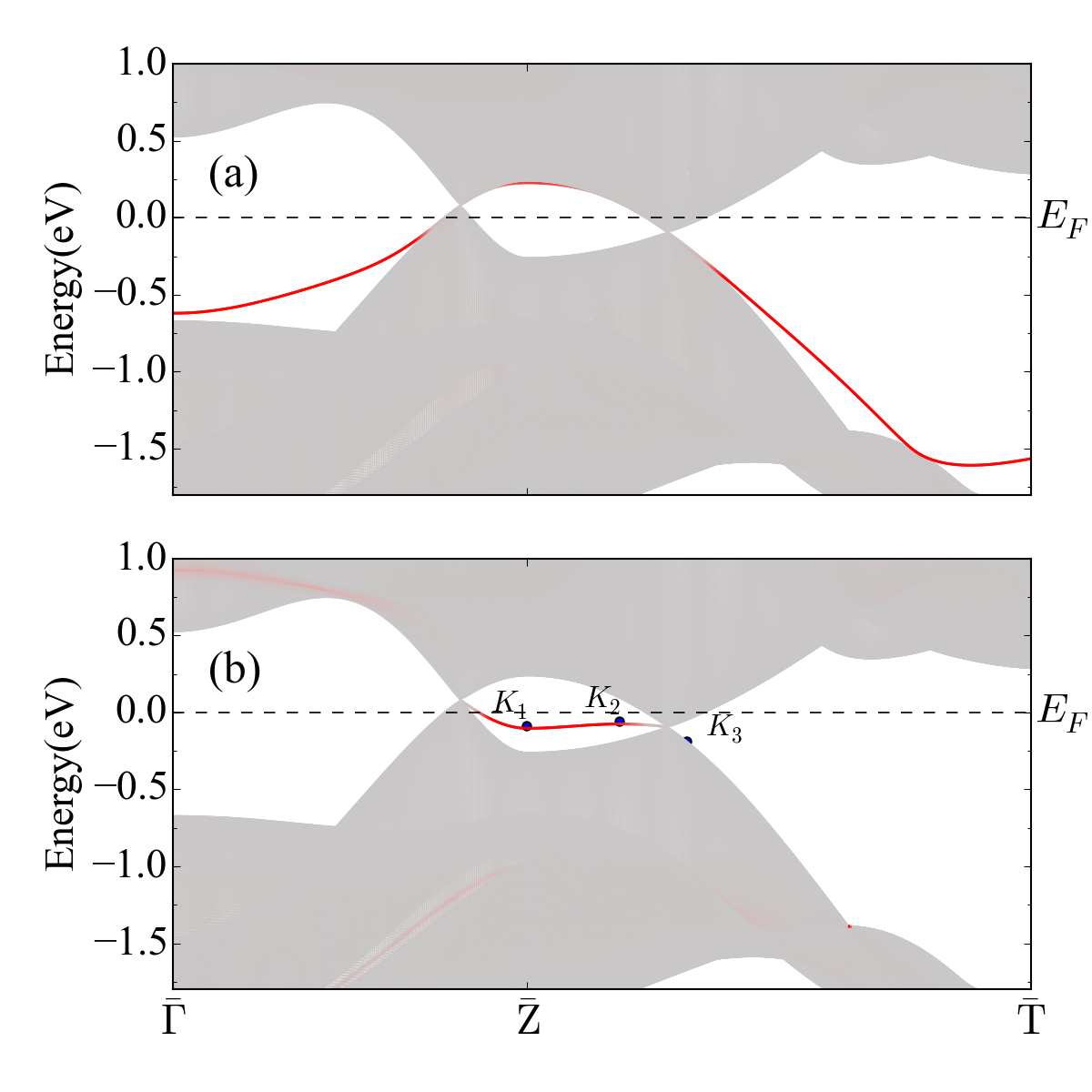}
\par\end{centering}
\protect\caption{\label{fig:surface_states}
(Color online) The surface states of compressed BP on 
the (a) zigzag type and (b) beard type surface in the (100) direction.
The surface states exist outside the node ring for the zigzag type
surface and inside the node ring  for the beard type surface.
}
\end{figure}

\begin{figure}[h]
\begin{centering}
\includegraphics[width=1.0\columnwidth]{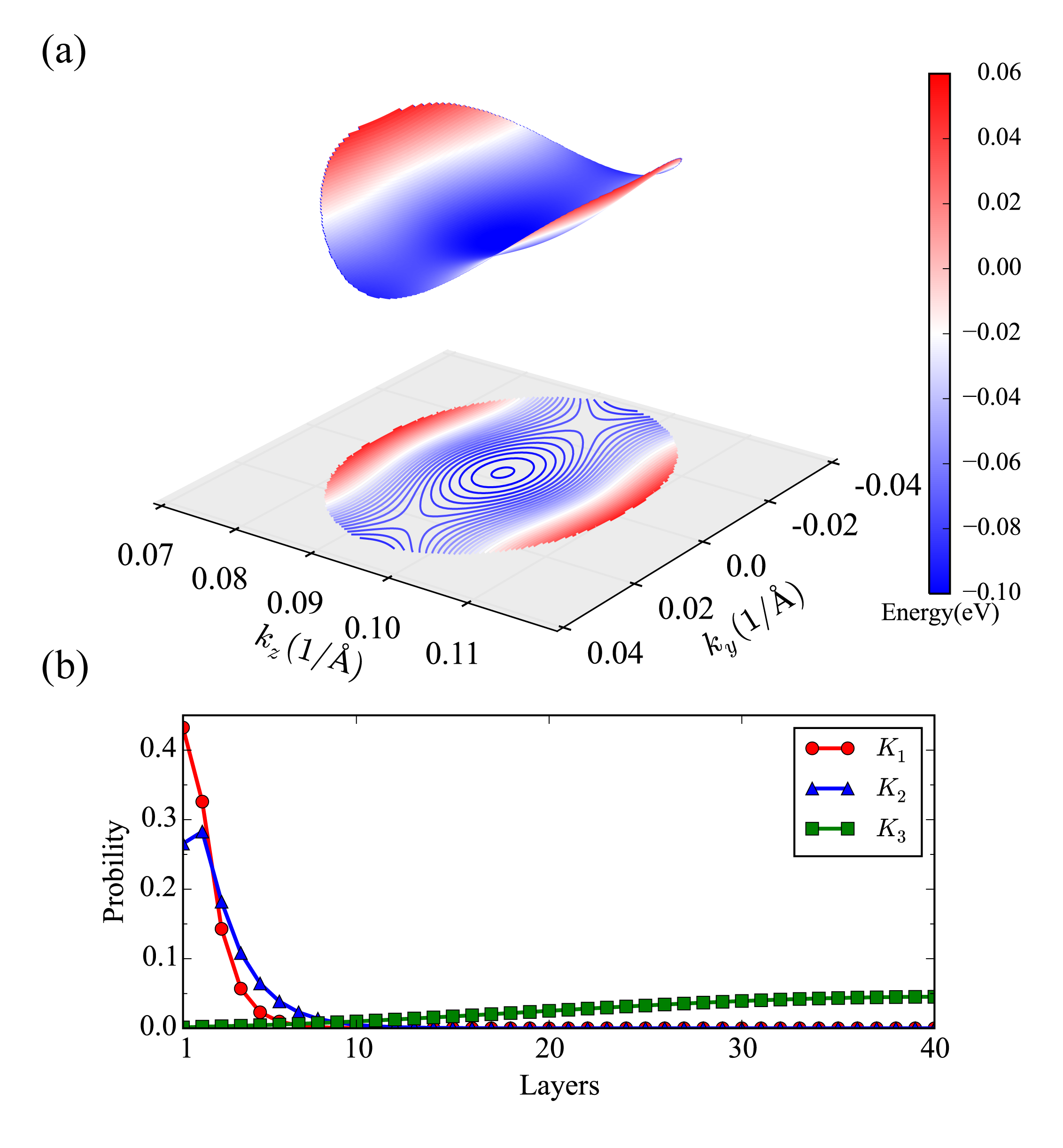}
\par\end{centering}
\caption{
(Color online)
(a) The side view of the surface state nested inside the node ring. 
(b) The real space distribution of wavefunction ($|\psi_{nk}(\mathbf{r})|^2$) at three different $k$-points ($K_1$, $K_2$ and $K_3$ which are indicted in Fig.~\ref{fig:surface_states}(b)) on a slab with 80 layers. The surface states are localized on the surface with penetration depth about 5 layers (5 $\mathrm{\AA}$ each layer). \label{fig:drumhead}}
\end{figure}

The band inversion and the node line in compressed BP indicates that a novel 
topologically nontrivial surface states can exist on the surface of bulk BP.
In order to calculate such surface states, 
we construct a tight-binding Hamiltonian from MLWF method and obtain a thick slab along the (100) direction.
There are two type of surface in the (100) direction: the zigzag type and the beard type as shown in Fig. (\ref{fig:CS_BZ}) (a, b).
For the zigzag surface, the calculated surface states are located outside the node ring as shown in 
Fig. (\ref{fig:surface_states}) (a).
We attach H atoms on the zigzag type surface to form the beard surface.
A nearly flat surface state exists inside the node ring as present in Fig.~\ref{fig:surface_states} (b).
The 2D structure  of the surface states 
are presented in Fig.~\ref{fig:drumhead} (a).
The nearly flat dispersion of the surface states are proposed as a route to achieving high-temperature superconductivity~\cite{PhysRevB.83.220503,Heikkila:2015vj}.
The real space distribution of wave functions at three special $k$ points as indicated in 
Fig.~\ref{fig:surface_states} (b) are presented in Fig.~\ref{fig:drumhead}(b).
At the center of the surface state, the wave function is localized on the surface region, with penetration depth about 5 layers (about 2.5 nm).
Moving away from the center, the distribution of wave function extend from surface to  bulk.

\subsection{Four-Bands Tight-Binding Model }\label{sec_TBmodel}

In this section, we analyze the properties of BP by performing a four-bands
tight-binding model. The tight-binding model is given as ~\cite{BP_TB_PRB_2014}
\begin{equation}
H=\sum_{i}\epsilon_{i}c_{i}^{\dagger}c_{i}+\sum_{i\ne j}t_{ij}c_{i}^{\dagger}c_{j},\label{eq:H_tb}
\end{equation}
where the summation $i$ runs over the lattice sites,  
$c_{i}^{\dagger}(c_{j})$ is the creation (annihilation) operator 
of electrons at site $i\;(j)$, $\epsilon_{i}$ is the on site energy
parameter, and $t_{ij}$ is the hopping parameter
between the $i$-th and $j$-th sites as shown in Fig. (\ref{fig:CS_BZ}) (a).
The fitted parameters read as $\epsilon=-1.1112$ eV, $t_{1}^{\parallel}=-1.3298$ eV, $t_{2}^{\parallel}=4.2265$ eV,
$t_{3}^{\parallel}=-0.3605$ eV, $t_{4}^{\parallel}=-0.1621$ eV, 
$t_{1}^{\bot}=0.5558$ eV,$t_{2}^{\bot}=0.2303$ eV.

By performing the Fourier transformation, we obtain the Hamiltonian
in momentum space as $H=\sum_{k}c^{\dagger}(k)H(k)c(k)$, with
\begin{equation}
H(k)=\epsilon+\left[\begin{array}{cccc}
0 & h_{12} & h_{13} & h_{14}\\
 & 0 & h_{23} & h_{24}\\
 &  & 0 & h_{34}\\
\dagger &  &  & 0
\end{array}\right],\label{eq:H(k)}
\end{equation}
where
\begin{eqnarray}
h_{12} & = & t_{1}^{\parallel}(1+e^{-ik\cdot a_{2}})+t_{3}^{\parallel}(e^{-ik\cdot a_{1}}+e^{-ik\cdot(a_{1}+a_{2})}),\nonumber \\
h_{13} & = & t_{4}^{\parallel}(1+e^{-ik\cdot a_{2}}+e^{-ik\cdot a_{1}}+e^{-ik\cdot(a_{1}+a_{2})})\nonumber \\
 &  & +t_{2}^{\bot}(e^{-ik\cdot a_{3}}+e^{-ik\cdot(a_{1}+a_{3})}),\nonumber \\
h_{14} & = & t_{2}^{\parallel}e^{-ik\cdot(a_{1}+a_{2})}\nonumber \\
 &  & +t_{1}^{\bot}(e^{-ik\cdot(a_{1}+a_{3})}+e^{-ik\cdot(a_{1}+a_{2}+a_{3})}),\\
h_{23} & = & t_{2}^{\parallel}+t_{1}^{\bot}(e^{-ik\cdot a_{3}}+e^{ik\cdot(a_{2}-a_{3})}),\nonumber \\
h_{24} & = & h_{13},\nonumber \\
h_{34} & = & h_{12},\label{eq:h_ij}
\end{eqnarray}
where $a_{1,2,3}$ are lattice axis and the basis functions $(\phi_1,\phi_2,\phi_3,\phi_4)$ located on the four P-atoms in the unit cell are sketched in Fig.~(\ref{fig:CS_BZ}) (a). The band structure calculated
by using the tight-binding model in comparison with the first-principle
bands are shown in Fig.~(\ref{fig:CS_BZ}) (d). One can see that the four-bands model
is good enough to describe the low energy band structure of BP. Near
the Fermi energy, both the valence and conduction band are well reproduced
within the energy region of $\pm 0.25$ eV. Beyond that region, the four-bands
model does not give a reliable description due to the limit number
of basis and also for the basis function with different symmetry compare to the p-orbits 
of phosphorus~\cite{BP_TB_PRB_2014}.

As propose in Ref.~\onlinecite{ryu_topological_2002} that the
surface states are directly related to the non-zero Berry phase of 
a one-dimensional systems. In the following section, we study 
the one-dimensional systems parameterized
by the in-plane momentum $k_{\parallel}=(k_{2},k_{3})$ 
for the zigzag and beard type surface and show 
its relation to the surface states for these two cases.

The Berry phase for the one-dimensional system is defined as
\begin{equation}
\theta(k_{\parallel})=\sum_{n\in occu.}\int_{k_{1}}dk_{1}A_{nn}^{1}(k_{\parallel})\label{eq:Berry_phase}
\end{equation}
where the summation $n$ runs over occupied states, $A^{1}(k_{\parallel})$
is the Berry connection matrix which is defined as $A_{mn}^{1}(k_{\parallel})=\langle u_{m}(\mathbf{k})|i\partial_{k_{1}}|u_{n}(\mathbf{k})\rangle$
with $m$ and $n$ are band indices for the occupied states. To calculate
the Berry phase numerically, we use the method proposed in 
Ref.~\onlinecite{yu_Z2_2011,Weng:2014sva,Weng_ADV:2015hh}, where 
the Berry connection $A(k_{\parallel})$ can be calculated discretely
by introduce the $F$ matrix which is defined as
\begin{equation}
F_{m,n}^{i,i+1}(k_{\parallel})=\langle m,k_{1,i},k_{\parallel}|n,k_{1,i+1},k_{\parallel}\rangle\approx e^{-iA_{m,n}^{i,i+1}(k_{\parallel})\delta k_{1}},\label{eq:F_matrix}
\end{equation}
where $i=0,...,N_{1}$ defines the discretized position along $k_{1}$
direction in the BZ and $\delta k=2\pi/N_{1}a_{1}$. We can define
the product of $F_{i,i+1}$s as
\begin{equation}
D(k_{\parallel})=F_{0,1}F_{1,2}...F_{N_{1}-2,N_{1}-1}F_{N_{1}-1,0}.\label{eq:D_matrix}
\end{equation}
Substitute Eq. (\ref{eq:F_matrix}) into above equation, we have that
\begin{eqnarray}
D(k_{\parallel}) & = & \prod_{i=0}^{N_{1}-1}F_{i,i+1}=\prod_{i=0}^{N_{1}-1}e^{-iA^{i,i+1}(k_{\parallel})\delta k}\nonumber \\
 & = & \left\{ Pexp\big[\int_{k_1}-iA^{1}(k_{\parallel})dk\big]\right\} .\label{eq:D_matrix_2}
\end{eqnarray}
Then the Berry phase can be obtained as 
\begin{equation}
\theta(k_{\parallel})=\Im\big\{log[det(D(k_{\parallel})]\big\}\;mod\;2\pi.\label{eq:Berry_phase_2}
\end{equation}
Here, the identical equation $det[e^{M}]=e^{trM}$ for square matrix
$M$ are used to obtain Eq.~(\ref{eq:Berry_phase_2}). 
The eigenstates for Hamiltonian~\ref{eq:H(k)} can be numerically
obtained and the Berry phase can be calculated following Eqs.
(\ref{eq:F_matrix}) to (\ref{eq:Berry_phase_2}). The calculated
Berry phase for the zigzag type surface equals $\pi$ for 
$k_\parallel$ outside the node ring, while it is zero for $k_\parallel$ inside the node ring.

For the beard type surface, the Hamiltonian matrix elements in Eq.~(\ref{eq:H(k)}) are listed as
\begin{eqnarray}
h_{12} & = & t_{2}^{\parallel}+t_{1}^{\bot}(e^{ik\cdot a_{3}}+e^{-ik\cdot(a_{2}-a_{3})}),\nonumber \\
h_{13} & = & t_{4}^{\parallel}(1+e^{-ik\cdot a_{2}}+e^{-ik\cdot a_{1}}+e^{-ik\cdot(a_{1}+a_{2})})\nonumber \\
 &  & +t_{2}^{\bot}(e^{-ik\cdot(a_{2}-a_{3})}+e^{-ik\cdot(a_{1}+a_{2}-a_{3})}),\nonumber \\
h_{14} & = & t_{1}^{\parallel}(e^{-ik\cdot a_{1}}+e^{-ik\cdot(a_{1}+a_{2})})+t_{3}^{\parallel}(1+e^{-ik\cdot a_{2}}),\nonumber \\
h_{23} & = & t_{1}^{\parallel}(1+e^{-ik\cdot a_{2}})+t_{3}^{\parallel}(e^{-ik\cdot a_{1}}+e^{-ik\cdot(a_{1}+a_{2})}),\nonumber \\
h_{24} & = & t_{4}^{\parallel}(1+e^{-ik\cdot a_{2}}+e^{-ik\cdot a_{1}}+e^{-ik\cdot(a_{1}+a_{2})})\nonumber \\
 &  & +t_{2}^{\bot}(e^{-ik\cdot a_{3}}+e^{-ik\cdot(a_{1}+a_{3})}),\nonumber \\
h_{34} & = & t_{2}^{\parallel}+t_{1}^{\bot}(e^{-ik\cdot a_{3}}+e^{ik\cdot(a_{2}-a_{3})}),\label{eq:h_in_beard}
\end{eqnarray}
where the basis is indicated in Fig.~(\ref{fig:CS_BZ}) (b). 
The calculated Berry phase for the beard surface is just on the contrary to the zigzag case, which is in agreement with the surface 
states as shown in Fig.~(\ref{fig:surface_states}).

\section{Conclusion}\label{sec_summary}

In summary, we propose that the 3D topological NLS states can be realized 
in the compressed bulk BP when SOC is ignored.
A closed node line is found near the Fermi energy, which is protected by time-reversal 
and inversion symmetry with band inverted in the bulk band structure.
The two dimensional surface states are studied in the (100) direction
terminated with zigzag and beard type surface.
The surface states are nested inside the closed node line on the
beard type surface. Its nearly flat energy dispersion is an ideal playground 
for many interaction induced nontrivial states, such as fractional topological insulator 
and high-temperature superconductivity.

\begin{acknowledgments}

This work was supported by the National Natural Science Foundation of China 
(No.11274359, No.11422428 and No.41574076), the 973 program of China 
(No.2011CBA00108 and No.2013CB921700), 
and the Strategic Priority Research Program (B) of the 
Chinese Academy of Sciences (No.XDB07020100).

\end{acknowledgments}

\bibliography{refs}

\end{document}